\documentstyle[aps,manuscript]{revtex}

\mathchardef\bepsilon="7F0F

\begin{document}

\draft

\title{Exact solution for a two-level 
atom in radiation fields\\ 
and the Freeman resonances}

\author{Dong-Sheng Guo$^{1,2}$, Yong-Shi Wu$^{3,4}$,
and Linn Van Woerkom$^{2}$}

\address{$^1$Permanent Address: Department of Physics,
Southern University and A{\rm \&}M College, \\
Baton Rouge, Louisiana 70813, USA}
\address{$^2$Department of Physics, Ohio State University,
Columbus, Ohio 43210, USA}
\address{$^{3}$Department of Physics, University of Utah,
Salt Lake City, Utah 84112, USA}
\address{$^{4}$Center for Advanced Study, Tsinghua University,
Beijing 100084, China}


\maketitle

\begin{abstract}
Using techniques of complex analysis in an algebraic approach,
we solve the wave equation for a two-level atom interacting
with a monochromatic light field exactly. A closed-form expression
for the quasi-energies is obtained, which shows that the
Bloch-Siegert shift is always finite, regardless of whether the
original or the shifted level spacing is an integral multiple
of the driving frequency, $\omega$. We also find that the
wave functions, though finite when the original level spacing
is an integral multiple of $\omega$, become divergent when
the intensity-dependent shifted energy spacing is an integral
multiple of the photon energy. This result provides, for the first
time in the literature, an ab-initio theoretical explanation for the
occurrence of the Freeman resonances observed in above-threshold
ionization experiments.
\end{abstract}



\section{Introduction}

The interaction between light and matter is a fundamental
problem in physics, whose study has led to the birth of
quantum theory about a century ago.
The two-level atom model was originally proposed by Einstein
\cite{Einstein} to study the transitions between two
energy levels of an atom interacting with light
(later especially with laser light).
However, despite many great
and significant progresses since then, an exact solution
remains elusive even for the simplest problem of a two-level
atom \cite{eberly} interacting with a classical or
quantum-mechanical light field.
Pursuing higher accuracy in describing a physical system is
always an ultimate goal for physicists.

Exact and analytic
expressions also provide new starting points for further
developments of theories in physics.
The exact quasi-energy levels and the wave functions 
obtained by solving the two-level atom model can
be used in the calculation of many
important physical quantities, such as the Rabi flopping
frequency and the inversion rate. Mathematically, an
approach that exactly solves this model may provide a
starting point for solving more complicated cases, such as
a driven N-level atom, as well as an atom in a multi-mode
laser field.

Since the pioneering work of Bloch and Siegert (BS) \cite{bs},
there have been many different approximate methods developed to
solve a two-level system driven by an external field. The rotating
wave approximation (RWA) is a widely used method. As is
well-known, this approximation is good only when the frequency of
the light field is ``near resonance'' but also not too close to
the resonance.  The word ``near'' means that the frequency of the
light field is near the original energy spacing of the two-level
atom. Many works have been devoted to go beyond the RWA method.
For example, Shirley\cite{shirley} applied Floquet's theorem and
perturbation method to solve the time-dependent Schr\"odinger
equation for a two-level system. And Cohen-Tannoudji {\it et al.}
\cite{cohen} used perturbation methods to solve a quantum-field
two-level system. Mittleman {\it et al.}\cite{mittle} have
recently derived the wave functions and quasi energies for a
two-level atom driven by a low-frequency strong laser pulse and
applied their result to emission spectra and high harmonic
generation (HHG). The low-frequency approximation adopted by
Mittleman {\it et al.} skips all higher resonances and can be
thought of as a limiting case opposite to the near-resonance
approximation. The Continued Fraction (CF) method, giving
recurrence relations for the Fourier coefficients of the wave
function, is also a commonly adopted method going beyond the RWA.
Swain \cite{swain}, Yeh and Stehle \cite{yeh}, Becker
\cite{becker}, and recently Feng \cite{feng} applied the CF method
to obtain approximate solutions. In continued fractions expressing
the wave function, the unknown quasi-energy is involved.
Approximations, used in evaluating the quasi-energy, make the
corresponding wave functions inaccurate. Due to the infinite order
of the algebraic equations satisfied by the quasi energy, the
exact value of quasi energy in a closed form has never been
derived.

In 1987, R. R. Freeman {\it et al.} found experimentally
\cite{free} that the above-threshold ionization (ATI) peaks broke
up into many small peaks when the laser pulses were short. The
appearance of the small peaks were interpreted as multiphoton
resonances between the ground state, say $5P_{3/2}$ for the
outermost shell electrons of xenon atoms, and Rydberg states with
a shifted energy level.  In the literature these resonances are
now called Freeman resonances and their appearance can be
phenomenologically explained by a.c. Stark-shifted multiphoton
resonances.

Such multiphoton resonances have been observed for years and
modelled theoretically using Floquet and numerical approaches
\cite{book}. Due to interactions with the radiation field, the
energy level spacing of a two-level atom acquires an intensity
dependent shift. In the theoretical literature ,the term ``near
resonance'' refers to the condition where the photon energy is
near the original energy spacing (pre-resonances, in the absence
of radiation). However, the fine structure of ATI peaks observed
by Freeman et al. \cite{free} can be well interpreted as Rydberg
state resonances occurring only when the shifted energy spacing is
equal to an integral multiple of the photon energy, the Freeman
resonances, in the presence of strong radiation fields. It is this
intensity-dependent shift in the resonance frequency that calls
for a fundamental explanation in the theory for a driven atom. 
Moreover, one is certainly tempted to know what happens when 
the light field is neither very near nor very far away from 
any resonance. Even in the so-called near-resonance case, 
calculations and analysis with higher accuracy are always 
desired. All of these require a solution to the two-level 
atom problem that is as exactly as possible.

In this paper, we attack the two-level atom problem in 
an algebraic approach. We start with a proof of the 
equivalence between a classical-field description and a 
corresponding quantum-field description for a two-level 
system in a driving field. Then, we recast the 
classical-field differential equations of motion into 
an infinite system of linear equations, with the 
energy determinant of infinite rank of the form of a
continuant \cite{muir}, i.e. having non-zero elements 
only on three major diagonals. We then directly evaluate 
the energy determinant, using techniques in complex 
analysis and the trick that breaks the relevant 
continuants into sub-continuants of a half infinite 
rank, which are further expressed as infinite series. 
In this way, a closed-form expression for the quasi
energies is obtained. 

Our solutions exhibit several interesting features: 
1) A simple cosine energy shift formula is derived, 
which naturally exhibits the Floquet quasi-energy 
feature. 2) It incorporates multiphoton effects, 
in particular all multi-photon resonances if there 
are any. 3) For the pre-resonance cases, when the 
original energy spacing is an integral multiple 
of the photon energy of the radiation field, both 
the shifted quasi-energies and the corresponding 
wave functions are finite. Therefore no singularity
(or resonance) really appears at the pre-resonances.
4) It shows theoretically the existence of Freeman 
resonances, namely the wave function has a singularity
when the intensity dependent energy spacing shifts 
to an integral multiple of the external photon energy.

Comparisons are made between our result and earlier 
results indeed shows agreement at the leading order. 
The higher order correctness of our results can be 
guaranteed and checked by the mathematical derivation 
process and also by the comparison with experimental
findings.

\section{Equations of Motion}

The goal of this paper is to solve the following
equations of motion which describes a two-level 
atom driven by a radiation field:
\begin{eqnarray}
({d \over d\tau} -iD\cos \tau
\sigma_x +i\Delta \sigma_z + iE I){\bf Y} =0.
\label{eqc}
\end{eqnarray}
In this equation, we have chosen $c=\hbar=1$,
and set the field frequency $\omega=1$. Here 
$2\Delta$ stands for the energy spacing in units 
of $\omega$ in the absence of the radiation 
field; we have introduced a dimensionless dipole 
moment $D$ for the interaction strength, with 
$D^2$ proportional to the laser beam intensity. 
The notation $E$ stands for the quasi-energy,
in uints of $\omega$, of the two-level atom 
in the classical radiation field. It can 
also be directly called the energy level, 
if one treats the radiation as a quantum 
field. ${\bf Y}=(Y_1,Y_2)^t$, and $\sigma_x$, 
$\sigma_z$ are Pauli matrices; $I$ is the 
$2\times 2$ unit matrix. 

This equation is usually derived from the quantum 
mechanical equations of motion for a two-level atom 
interacting with a classical, single-frequency mode 
in the {\it dipole} approximation. Below we will 
show that it can also be derived from the quantum 
field approach in the large photon number (LPN)
limit without any other approximation. For this 
reason, we regard the equations of motion 
(\ref{eqc}) as an exact description of a driven 
two-level system when the driving field is a 
classical field.

Let us start from the equations of motion for a
two-level atom interacting with a single-mode
quantun field:   
\begin{eqnarray}
 [\Delta \sigma_z+\omega N
+|e|g\sigma_x(a+a^{\dag})]|\phi\rangle
=-E|\phi\rangle,
\label{eqq}
\end{eqnarray}
which is equivalent to the one in Cohen-Tannoudji 
{\it et al.}'s paper \cite{cohen}, if we set 
$2\Delta=\omega_0, {1\over2}{\bf \sigma} ={\bf J},
g={\lambda\over4|e|}, E=-E'$.

Now, we introduce a new basis 
$\langle y|\equiv\sum_n y^n\langle n|$ with 
$\langle n|$ being Fock states:
\begin{eqnarray}
|\phi\rangle \to \phi(y)\equiv
\sum_ny^n\langle n|\phi\rangle.
\label{trans}
\end{eqnarray}
Then, in the large-photon-number (LPN) limit,
the equations of motion (\ref{eqq}) becomes 
\begin{eqnarray}
\sum_n [\Delta \sigma_zy^n+ny^n\omega
+|e|\Lambda\sigma_x(y^{n-1}+y^{n+1})]\phi_n 
=-E\sum_n y^n\phi_n,
\label{eq1}
\end{eqnarray}
where
\begin{eqnarray}
\phi_n\equiv \langle n|\phi\rangle, 
\quad \Lambda\equiv g \sqrt{n}. 
\label{phin}
\end{eqnarray}

We further rewrite the above equation as
\begin{eqnarray}
[\Delta \sigma_z+\omega y{d\over dy}
+|e|\Lambda\sigma_x(y^{-1}+y)]\phi(y) 
=-E\phi(y). 
\label{eq1a}
\end{eqnarray}
Letting
\begin{eqnarray}
2|e|\Lambda\equiv D\omega, \quad 
y\equiv -e^{i\omega\tau}, \quad
\omega=1, \label{tau}
\end{eqnarray}
we obtain eq. (\ref{eqc}). Starting with 
eq. (\ref{eqc}) and going backward through
the proof, with mapping $y^n$ to Fock state
$|n\rangle$ and resuming the commutation 
relation between $y$ and $y^{-1}$, we 
can recover eq. (\ref{eqq}). To recover 
$\omega$, one should just take 
$\Delta \to \Delta/\omega$ and 
$E \to E/\omega$, with $D$ remaining 
dimensionless. 

\section{Solving the Algebraic Equations of Motion}

Write the solutions in the case of
$\Delta\ne 0$ in the form:
\begin{eqnarray}
{\bf Y}= C_1 (\tau) \bar {\bf Y}^{(1)}
+ C_2 (\tau) \bar {\bf Y}^{(2)},
\label{e14}
\end{eqnarray}
where $\bar {\bf Y}^{(1)}$ and
$\bar {\bf Y}^{(2)}$ are two
linearly independent solutions of eq.
(\ref{eqc}) with $\Delta=0=E$:
\begin{eqnarray}
\bar{\bf Y}^{(1)}(\tau)
=e^{-i D\sin \tau}{1\over\sqrt{2}}(1,-1)^t,\quad
\bar {\bf Y}^{(2)}(\tau)
=e^{i D\sin \tau}{1\over\sqrt{2}}(1,1)^t.
\label{e15}
\end{eqnarray}
Here the superscript $t$ means transposition.
These solutions are ortho-normal:
$\bar{\bf Y}^{(i)\dag}\bar{\bf Y}^{(j)}
=\delta_{ij}$, and satisfy
$$\bar {\bf Y}^{(1)\dag} \sigma_z
\bar {\bf Y}^{(1)}=
\bar {\bf Y}^{(2)\dag} \sigma_z
\bar {\bf Y}^{(2)}= 0,  $$
$$\bar {\bf Y}^{(1)\dag} \sigma_z
\bar {\bf Y}^{(2)}
= (\bar {\bf Y}^{(2)\dag} \sigma_z
\bar {\bf Y}^{(1)})^*
=e^{i 2D\sin \tau}$$

The resulting differential equations are
\begin{eqnarray}
i{d\over d\tau} C_1(\tau) -
\Delta C_2(\tau) e^{i2D\sin\tau}
- EC_1(\tau)=0 \nonumber\\
i{d\over d\tau}C_2(\tau) 
- \Delta C_1(\tau) e^{-i2D\sin\tau} 
- EC_2(\tau)=0. \label{e16}
\end{eqnarray}
Using the following expansions,
\begin{eqnarray}
C_1(\tau) = \sum_s C_{1s} e^{-is\tau},
\qquad
C_2(\tau) = \sum_s C_{2s} e^{-is\tau},
\label{e17}
\end{eqnarray}
Equations (\ref{e16}) can be transformed 
into a set of linear equations:
\begin{eqnarray}
&&s C_{1s} - \Delta \sum_t C_{2t}
{\rm J}_{t-s}(2D) - E C_{1s}=0, \nonumber \\
&&s C_{2s} - \Delta \sum_t C_{1t}
{\rm J}_{t-s}(-2D) - E C_{2s}=0,
\label{e18}
\end{eqnarray}
where ${\rm J}_n(x) $ are ordinary Bessel 
functions, $s$ and $t$ are integers running 
from $-\infty$ to $\infty$. However, each 
of the equations involves an infinite sum 
of terms, so they still look awful. To
simplify them, we use the Bessel functions 
to construct a transformation:
\begin{equation}
A_q \equiv \sum_s C_{1s}
{\rm J}_{s-q}(-2D).
\end{equation}
The inverse transformation is
\begin{eqnarray}
C_{1s}=\sum_q A_q {\rm J}_{q-s}(2D).
\label{e19}
\end{eqnarray}
Using the inverse transformation to express
eq. (\ref{e18}), and the recurrence relations 
for the Bessel functions $n{\rm J}_n(2D)
=D [ {\rm J}_{n-1}(2D)+{\rm J}_{n+1}(2D)]$,
we obtain
\begin{eqnarray}
&&
\Delta C_{2s}=-D(A_{s+1}+A_{s-1})+(s-E)A_s,
\nonumber\\
&&
C_{2s}={\Delta\over s-E} A_s.
\label{e19'}
\end{eqnarray}

So finally the new variables $A_s$ satisfy 
a set of simple linear equations
\begin{eqnarray}
{\frac{D(E-s)}{(E-s)^2-\Delta^2}} 
(A_{s+1}+A_{s-1})+ A_s=0,
\label{A-recur}
\end{eqnarray}
with $s=\cdots, -2,-1,0,1,2,\cdots$. Each
equation now involves only three terms. 
This success of simplification is crucial 
to our subsequent treatments.

\section{Infinite Determinant and Quasi-Energies}

For a nontrivial solution to eqs. (\ref{A-recur})
to exist, the quasi energy $E$ has to be such
that the following infinite determinant vanishes:
\begin{eqnarray}
 \det(E) =
\left|\matrix{\cdots
&\cdots &\cdots  &\cdots
&\cdots &\cdots &\cdots  \cr
\cdots
& \alpha_{-2} & \beta_{-2}(E)
&0 &0 &0 &\cdots \cr
\cdots
& \gamma_{-1}(E)
& \alpha_{-1} & \beta_{-1}(E)
&0 &0 &\cdots \cr
\cdots  &0
& \gamma_{0}(E) & \alpha_{0}
& \beta_{0}(E)&0 &\cdots\cr
\cdots &0 &0
& \gamma_{1}(E)
& \alpha_{1} & \beta_{1}(E)
&\cdots \cr
\cdots  &0 &0 &0
& \gamma_{2}(E)
& \alpha_{2} &\cdots \cr
\cdots &\cdots &\cdots &\cdots
&\cdots &\cdots &\cdots\cr}
\right|,
\label{det-E}
\end{eqnarray}
with
\begin{eqnarray}
\alpha_{s}=1,\quad \beta_{s}(E) = \gamma_{s+1}(E)=
\frac{D(E-s)}{(E-s)^2-\Delta^2}.
\label{abc}
\end{eqnarray}
A determinant of this type is called tri-diagonal, 
or it is called a continuant \cite{muir}.

In the previous literature, this kind of infinite 
determinants were evaluated by using various 
approximations, e.g., the power series expansion 
in $D$ \cite{feng}. Here we will evaluate this
infinite determinant exactly, using techniques in 
complex analysis. The key observation is that the 
tri-diagonal infinite determinant in eq. (\ref{det-E}) 
is absolutely convergent, since the infinite sum, 
$\sum_s \beta_s \gamma_{s+1}$, is absolutely 
convergent. (See, for example, the classical
treatise \cite{whitt}.)


Therefore, if we regard the energy $E$ as a complex 
variable, then the infinite determinant (\ref{det-E}) 
defines an analytic function on the complex-$E$ plane. 
Actually it is a meromorphic function of $E$ which 
has two groups of poles at $E=\pm (\Delta+s)$, 
$(s=0\pm1,\pm2,\cdots)$.

When $2\Delta=n$ for integer $n \geq 1$, we call the 
case as {\it pre-resonance}. In the non-pre-resonant 
case where $2\Delta\neq n$, the poles of $\det(E)$ 
are all simple poles. Since $\beta_s(E)$ and 
$\gamma_s(E)$ are periodic functions of $E$ with
period unity. The poles in each group, 
$E=\pm (\Delta+s)$ respectively, are equally 
spaced and have the same residue. The residues,
\begin{eqnarray}
R_{+}= \lim_{E-\Delta\pm n \to 0} 
(E -\Delta \pm n)\det(E), \qquad
R_{-}=\lim_{E+\Delta \pm n \to 0} 
(E +\Delta \pm n)\det(E),
\label{cond}
\end{eqnarray}
are independent of $n$. Furthermore, it can be  
directly verified that the residues $R_{+}$ and 
$R_{-}$ differ from each other only by a sign:
\begin{equation}
R_{+} (D,\Delta)=- R_{-} (D,\Delta).
\label{N-rel}
\end{equation}

Each group of poles with the same residue on the 
real $E$ axis of the complex-$E$ plane suggests 
a cotangent function. We also observe that 
$\det(E=i\infty)=1.$ Thus the exact value of the
determinant (\ref{det-E}) can only be the 
following function:
\begin{eqnarray}
\det(E)&&=1+R_{+}\sum_{n=-\infty}^{\infty}
          ({1\over E-\Delta+n}-{1\over E+\Delta+n})\nonumber\\
       &&=1+\pi R_{+} \cot [ \pi(E-\Delta) ]
         -\pi R_{+} \cot [ \pi(E+\Delta) ] \nonumber \\
       &&=1+\pi R_{+} {\sin(2\pi\Delta)
           \over\sin [ \pi(E-\Delta) ]\sin [ \pi(E+\Delta) ]} .
\label{cot}
\end{eqnarray}
By the uniqueness theorem in complex analysis, the right side of
eq. (\ref{cot}) and that of eq. (\ref{det-E}) agree with each
other on the whole $E$-plane.

When $2\Delta$ is a positive integer, the two groups of simple
poles are merged to become double poles. When $\Delta=0$, the
poles at $E= integer$ still remain as a first-order or zeroth-
order ones. (The details will be discussed in Appendix.)

Thus we are able to put the characteristic equation, 
$\det(E)=0$, into an exact and closed form:
\begin{eqnarray}
    \sin [ \pi(E-\Delta) ]\sin [ \pi(E+\Delta) ]
    =-\pi R_{+} \sin(2\pi\Delta) \; .
\label{e37}
\end{eqnarray}
Solving the above equation, with inclusion of the 
pre-resonance case, we obtain a cosine function of 
quasi-energy:
\begin{eqnarray}
&&
\cos(2\pi E)
= \cos(2\pi\Delta)+2\pi R_{+} \sin(2\pi\Delta),
\quad\quad(2\Delta\ne n)
\nonumber\\
&& \cos(2\pi E) =\cos(2\pi\Delta)+(-1)^n2\pi^2 r_{n}, 
\quad\quad (2\Delta=n), \label{cosshft}
\end{eqnarray}
where $r_n$ are residues of $R_{+}$ as a function 
of $2\Delta$, defined by the following limiting 
processes:
\begin{eqnarray}
\lim_{2\Delta\to n}R_{+} \sin(2\pi\Delta)
=(-1)^n\lim_{2\Delta\to n} R_{+} 
\pi(2\Delta- n)=(-1)^n\pi r_{n}.
\label{residue}
\end{eqnarray}
We note that $r_0=0$ for $n=0$. (The detailed proof 
is presented in the Appendix.)

This expression (\ref{cosshft}) for energy shift has 
a unique advantage: the Floquet condition is 
automatically satisfied in view of the cosine function.

The next main job is to evaluate the factor $R_{+}$. In the
context below, for writing convenience, we use finite determinant
notations to express the infinite determinants. We find
\begin{eqnarray}
&&R_{+}= \lim_{E\to\Delta} (E -\Delta) \det(E)\nonumber \\
&&=\left|\matrix{ 1 &{D(\Delta+3)\over3(2\Delta+3) } &\cdot
&\cdot &\cdot &\cdot &\cdot \cr {D(\Delta+2)\over2(2\Delta+2) }&
1& {D(\Delta+2)\over2(2\Delta+2) } &\cdot &\cdot &\cdot &\cdot \cr
\cdot & {D(\Delta+1)\over(2\Delta+1) }& 1&
{D(\Delta+1)\over(2\Delta+1) } &\cdot &\cdot &\cdot \cr \cdot
&\cdot & {D\over 2} & 0 &{D\over 2} &\cdot &\cdot \cr \cdot &\cdot
&\cdot & {D(\Delta-1)\over-1\cdot(2\Delta-1) } & 1 &
{D(\Delta-1)\over-1\cdot(2\Delta-1) } &\cdot \cr \cdot  &\cdot
&\cdot &\cdot & {D(\Delta-2)\over-2(2\Delta-2) } & 1
&{D(\Delta-2)\over-2(2\Delta-2) }\cr \cdot &\cdot &\cdot &\cdot
&\cdot &{D(\Delta-3)\over-3(2\Delta-3) } &1 \cr} \right|.
\label{R_{+}}
\end{eqnarray}

The following lemma is useful to determine
the residues of $R_{+}$: 

{\it LEMMA}: For any integer $n$, the residues of
$\det(E)$ with $2\Delta=n$ at the second order 
poles $E=l+n/2$, where $l$ is an arbitrary integer, 
are equal to the residue of $R_{+}(2\Delta)$
at the first order pole $2\Delta=n$; i.e.,
\begin{eqnarray}
\lim_{E\to (l+n/2)}[E-(l+n/2)]^2\det(E)
\vert_{2\Delta=n} =r_n,
\label{lemma}
\end{eqnarray}
where $r_n$ are residue of the function 
$R_{+}(2\Delta)$ at the pole $2\Delta=n$.

{\it Proof}:
From eqs. (\ref{cot}) and (\ref{residue})
\begin{eqnarray}
\det(E)\vert_{2\Delta=n} =1+{(-1)^n\pi^2 r_{n}\over \sin [
\pi(E-\Delta) ]\sin [ \pi(E+\Delta) ]} \vert_{2\Delta=n} =1+
{\pi^2 r_{n}\over \sin^2[ \pi(E-l-n/2) ]}, \label{lemprf}
\end{eqnarray}
we can see that when $E\to (l+n/2)$, the function
$\pi^2/\sin^2[\pi(E-l-n/2) ]$ behaves like
a second order pole $[E-(l+n/2)]^{-2}$.

In the $n=0$ case, the above proof still holds, since
$R_{+}\vert_{2\Delta=0}=0$ (see appendix) and $r_0=0$.
{\it QED}

From this lemma we learn that the function of 
$R_{+}(2\Delta)$ can only have first-order poles 
at non-zero integers. Thus, $R_{+}$ can be 
expressed as
\begin{eqnarray}
R_{+}=R_{+}^{\infty}
+\sum_{n=-\infty}^{\infty}{r_n\over2\Delta-n}
=\sum_{n=1}^{\infty}{4\Delta \over4\Delta^2-n^2}r_n,
\label{laurent}
\end{eqnarray}
where (a) $r_n=r_{-n}$, (b) $r_0=0$, and (c) 
$R_{+}^{\infty}=0$ have been used. The proofs 
for (b) and (c) are given in Appendix. The 
proof for (a) is the following: Using eq. 
(\ref{lemma}) to express $r_{-n}$, we verify that
the values of the two factors on the left hand side
of this equation do not change with changing $n\to -n$.
That the second factor does not change can be seen from
eqs. (\ref{cot}) and (\ref{N-rel}) while the first one
can be seen with substituting $l$ in $l-n/2$ by $l'+n$ 
since both $l$ and $l'$ can be arbitrary integers.

Further evaluation of $R_{+}$ is based on the 
evaluation of the residues $r_n$. In the following, 
we list few low-photon-number residues as examples:
\begin{eqnarray}
r_{1}
=\left|\matrix{ 1 &{{7\over2}D\over3\cdot4 } &\cdot  &\cdot
&\cdot &\cdot &\cdot &\cdot \cr
{{5\over2}D\over2\cdot3 }& 1& {{5\over2}D\over2\cdot3 }
&\cdot &\cdot &\cdot &\cdot &\cdot \cr
\cdot
& {{3\over2}D\over1\cdot2 }& 1& {{3\over2}D\over1\cdot2 }
&\cdot &\cdot &\cdot &\cdot \cr
\cdot  &\cdot
& {D\over 2} & 0 &{D\over 2}
&\cdot &\cdot &\cdot \cr
\cdot &\cdot &\cdot
& {D\over2 } & 0 & {D\over2}
&\cdot &\cdot \cr
\cdot  &\cdot &\cdot &\cdot
& -{{3\over2}D\over2\cdot1 }
& 1 &-{{3\over2}D\over2\cdot1 } &\cdot\cr
\cdot &\cdot
&\cdot &\cdot &\cdot &-{{5\over2}D\over3\cdot2 } &1
&-{{5\over2}D\over3\cdot2 }\cr
\cdot &\cdot &\cdot
&\cdot &\cdot &\cdot &-{{7\over2}D\over4\cdot3 } &1
\cr} \right|,
\label{r1}
\end{eqnarray}

\begin{eqnarray}
r_{2}
=\left|\matrix{ 1 &{4D\over3\cdot5 } &\cdot  &\cdot
&\cdot &\cdot &\cdot&\cdot &\cdot \cr
{3D\over2\cdot4 }& 1& {3D\over2\cdot4 }
&\cdot &\cdot &\cdot &\cdot&\cdot &\cdot \cr
\cdot
& {2D\over1\cdot3 }& 1& {2D\over1\cdot3 }
&\cdot &\cdot &\cdot &\cdot &\cdot\cr
\cdot  &\cdot
& {D\over 2} & 0 &{D\over 2}
&\cdot &\cdot &\cdot &\cdot\cr
\cdot &\cdot &\cdot
& 0 & 1 & 0 &\cdot &\cdot &\cdot\cr
\cdot  &\cdot &\cdot &\cdot
& {D\over2 } & 0 & {D\over2} &\cdot &\cdot\cr
\cdot &\cdot
&\cdot &\cdot &\cdot &-{2D\over3\cdot1 } &1&
-{2D\over3\cdot1 }&\cdot \cr
\cdot &\cdot&\cdot
&\cdot &\cdot &\cdot &-{3D\over4\cdot2 } &1&
-{3D\over4\cdot2 } \cr
\cdot &\cdot&\cdot&\cdot
&\cdot &\cdot &\cdot &-{4D\over5\cdot3 } &1\cr}
\right|,
\label{r2}
\end{eqnarray}

\begin{eqnarray}
r_3
=\left|\matrix{
1&{D{7\over2}\over5\cdot2}
&\cdot &\cdot&\cdot &\cdot &\cdot &\cdot\cr
{D{5\over2}\over4\cdot1}
& 1 &{D{5\over2}\over4\cdot1}
&\cdot &\cdot &\cdot &\cdot&\cdot\cr
\cdot &{D\over2 }
& 0&{D\over2}
&\cdot &\cdot &\cdot&\cdot\cr
\cdot &\cdot
& {D{1\over2}\over2\cdot(-1)} & 1
&{D{1\over2}\over2\cdot(-1)}
&\cdot &\cdot&\cdot\cr
\cdot  &\cdot &\cdot
& {D(-{1\over2})\over1\cdot(-2)}
& 1 &{D(-{1\over2})\over1\cdot(-2)}&\cdot&\cdot\cr
\cdot &\cdot &\cdot&\cdot
& {D\over2}
& 0 & {D\over2}&\cdot\cr
\cdot &\cdot &\cdot &\cdot &\cdot
& {D(-{5\over2})\over(-1)(-4)}
& 1 & {D(-{5\over2})\over(-1)(-4)} \cr
\cdot &\cdot &\cdot&\cdot &\cdot &\cdot
& {D(-{7\over2})\over(-2)(-5)}& 1  \cr
}
\right|,
\label{r3}
\end{eqnarray}
and
\begin{eqnarray}
r_4=
\left|\matrix{
1&{D4\over6\cdot2} &\cdot  &\cdot &\cdot  &\cdot
&\cdot &\cdot &\cdot \cr
{D3\over5\cdot1} & 1&{D3\over5\cdot1}
&\cdot  &\cdot &\cdot
&\cdot &\cdot &\cdot \cr
\cdot & {D\over2}&0&{D\over2}
&\cdot &\cdot &\cdot &\cdot  &\cdot \cr
\cdot  &\cdot & {D1\over3\cdot(-1) }
& 1& {D1\over3\cdot(-1) }
&\cdot &\cdot &\cdot &\cdot \cr
\cdot &\cdot &\cdot
& 0 & 1
&0&\cdot &\cdot &\cdot \cr
\cdot &\cdot  &\cdot &\cdot
& {D(-1)\over1\cdot(-3)}
& 1 &{D(-1)\over1\cdot(-3)}&\cdot &\cdot
 \cr
\cdot  &\cdot  &\cdot &\cdot &\cdot & {D\over 2} & 0 & {D\over 2}
&\cdot \cr \cdot &\cdot &\cdot&\cdot &\cdot &\cdot
&{D(-3)\over(-1)\cdot(-5)} & 1 &{D(-3)\over(-1)\cdot(-5)}\cr \cdot
&\cdot &\cdot&\cdot &\cdot &\cdot &\cdot
&{D(-4)\over(-2)\cdot(-6)} & 1 \cr } \right| \label{r4}.
\end{eqnarray}

For general $n$, $r_n$ can be expressed as an determinant 
of infinite rank
\begin{eqnarray}
r_n
=\left|\matrix{ 1 &{D({n\over2}+3)\over3(n+3) } &\cdot  &\cdot
&\cdot &\cdot &\cdot \cr
{D({n\over2}+2)\over2(n+2) }& 1& {D({n\over2}+2)\over2(n+2) }
&\cdot &\cdot &\cdot &\cdot \cr
\cdot
& {D({n\over2}+1)\over1(n+1) }& 1& {D({n\over2}+1)\over1(n+1) }
&\cdot &\cdot &\cdot \cr
\cdot  &\cdot
& {D\over 2} & 0 &{D\over 2}
&\cdot &\cdot \cr
\cdot &\cdot &\cdot
& {D({n\over2}-1)\over-1(n-1) } & 1 &
{D({n\over2}-1)\over-1(n-1) }
&\cdot \cr
\cdot  &\cdot &\cdot &\cdot
& {D({n\over2}-2)\over-2(n-2) }
& 1 &{D({n\over2}-2)\over-2(n-2) }\cr
\cdot &\cdot
&\cdot &\cdot &\cdot &{D({n\over2}-3)\over-3(n-3) } &1 \cr
\cdot &\cdot &\cdot \cr
\cdot &\cdot &\cdot \cr
\cdot &\cdot &\cdot \cr
 1 &{D({n\over2}+3)\over3(n+3) } &\cdot  &\cdot
&\cdot &\cdot &\cdot \cr
{D({n\over2}+2)\over2(n+2) }& 1& {D({n\over2}+2)\over2(n+2) }
&\cdot &\cdot &\cdot &\cdot \cr
\cdot
& {D({n\over2}+1)\over1(n+1) }& 1& {D({n\over2}+1)\over1(n+1) }
&\cdot &\cdot &\cdot \cr
\cdot  &\cdot
& {D\over 2} & 0 &{D\over 2}
&\cdot &\cdot \cr
\cdot &\cdot &\cdot
& {D({n\over2}-1)\over-1(n-1) } & 1 &
{D({n\over2}-1)\over-1(n-1) }
&\cdot \cr
\cdot  &\cdot &\cdot &\cdot
& {D({n\over2}-2)\over-2(n-2) }
& 1 &{D({n\over2}-2)\over-2(n-2) }\cr
\cdot &\cdot
&\cdot &\cdot &\cdot &{D({n\over2}-3)\over-3(n-3) } &1 \cr}
\right|.
\label{rn}
\end{eqnarray}
In this notation, the lower part in the expression is the
continuation of the upper part in the direction of the main
diagonal.

We have been able to work out the first a few terms
for the residues in the dipole expansion. Here we only 
cite the results:
\begin{eqnarray}
&&
r_1=-{1\over4}D^2
+\left({\pi^2\over48}+{1\over64}\right)D^4+...
\nonumber\\
&&
r_2=-{1\over9}D^4+{13\over162}D^6+...
\nonumber\\
&& r_n=-{1\over4}\left({n\over n^2-1}\right)^2D^4
+...\quad\quad(n\ne 1), \label{dipole}
\end{eqnarray}
The numerical coefficients given here are all exact.

Thus, the expression for $R_{+}$ up to the $D^4$ term is
\begin{eqnarray}
R_{+}=\left[-{D^2\over4}
+D^4\left({\pi^2\over 48}+{1\over 64}\right)+...\right]
{4\Delta\over4\Delta^2-1}+\sum_{n=2}^{\infty}\left[
-{D^4\over4}\left({n\over n^2-1}\right)^2
+...\right]{4\Delta \over4\Delta^2-n^2}.
\label{R_{+}a}
\end{eqnarray}
The explicit form of $R_{+}$ can be used to evaluate the energy
shift. In the pre-resonance case
\begin{eqnarray}
&&
\cos(2\pi E)
=-1+{\pi^2\over2}D^2
-\pi^2\left({\pi^2\over24}+{1\over32}\right)D^4
+...,\quad\quad (2\Delta=1)
\nonumber\\
&& \cos(2\pi E) =(-1)^n -(-1)^n{\pi^2\over2} \left({n\over
n^2-1}\right)^2D^4+... \quad\quad (2\Delta=n\ne1).
\label{cosshft'}
\end{eqnarray}

There are many exact ways to express $E$ as an exact
function of $\Delta$. Following are the suggested ones:
\begin{eqnarray}
&&
 E= {1\over\pi}\cos^{-1}
\sqrt{\cos^2(\pi\Delta)+\pi R_{+} \sin(2\pi\Delta)},
\quad\quad(2\Delta\ne n)
\nonumber\\
&&
 E={1\over\pi}\cos^{-1}(\pi \sqrt {-r_{2k+1}})
\quad\quad (2\Delta=2k+1), \label{arccos}
\end{eqnarray}
and
\begin{eqnarray}
&&
 E= {1\over\pi}\sin^{-1}
\sqrt{\sin^2(\pi\Delta)-\pi R_{+} \sin(2\pi\Delta)},
\quad\quad(2\Delta\ne n)
\nonumber\\
&&
 E={1\over\pi}\sin^{-1}(\pi \sqrt {-r_{2k}})
\quad\quad (2\Delta=2k). \label{arcsin}
\end{eqnarray}
The right side of the equations can all be added with 
an integer, due to the Floquet feature for quasi-energy.

The Bloch-Siegert shift, defined as $E_{BS}=2E-2\Delta$, 
has now the exact expressions
\begin{eqnarray}
&&
 E_{BS}={2\over\pi}\cos^{-1}
\sqrt{\cos^2(\pi\Delta)+\pi R_{+} \sin(2\pi\Delta)}
-2\Delta,
\quad\quad(2\Delta\ne n)
\nonumber\\
&&
 E_{BS}={2\over\pi}\cos^{-1}(\pi \sqrt {-r_{2k+1}})
-2\Delta\quad\quad (2\Delta=2k+1), \label{bsarccos}
\end{eqnarray}
and
\begin{eqnarray}
&&
 E_{BS}={2\over\pi}\sin^{-1}
\sqrt{\sin^2(\pi\Delta)-\pi R_{+} \sin(2\pi\Delta)}
-2\Delta,\quad\quad(2\Delta\ne n)
\nonumber\\
&&
 E_{BS}={2\over\pi}\sin^{-1}(\pi \sqrt {-r_{2k}})
-2\Delta\quad\quad (2\Delta=2k). \label{bsarcsin}
\end{eqnarray}

\section{The Wave Functions}

The equation (\ref{A-recur})
can be further written as \cite{chien}
\begin{eqnarray}
&& {\rm for}\quad s > 0, \quad {A_{s}
\over A_{s-1}}={-1 \over\displaystyle
{(E-s)^2 - \Delta^2 \over\displaystyle
D(E-s)} + {A_{s+1} \over\displaystyle  A_{s}}},
\label{rel-1}\\
&& {\rm for}\quad s < 0, \quad {A_{s}
\over A_{s+1}}={-1 \over\displaystyle
{(E-s)^2 - \Delta^2 \over\displaystyle
D(E-s)} + {A_{s-1} \over\displaystyle  A_{s}}},
\label{rel-2} \\
&& {\rm for}\quad s = 0, \quad
{A_{-1}\over A_0} + { E^2 - \Delta^2 \over DE}
+ {A_1\over A_0 } = 0.
\label{rel-3}
\end{eqnarray}

By iterating the first relation, we express
$A_1/A_0$ as a continued fraction:
\begin{eqnarray}
{A_{1} \over A_{0}}=-{1 \over\displaystyle
{(E-1)^2 - \Delta^2 \over\displaystyle
D(E-1)} - {1 \over\displaystyle
{(E-2)^2 - \Delta^2 \over\displaystyle
D(E-2)} - ...}}.
\label{e25}
\end{eqnarray}
Similarly by iterating the second relation, we have
\begin{eqnarray}
{A_{-1} \over A_{0}}=-{1 \over\displaystyle
{(E+1)^2 - \Delta^2 \over\displaystyle
D(E+1)} - {1 \over\displaystyle
{(E+2)^2 - \Delta^2 \over\displaystyle
D(E+2)} - ...}}.
\label{e26}
\end{eqnarray}

Putting back all the transformations made, we obtain
\begin{eqnarray}
&&{\bf Y}^\mp= \sum_s \sum_q A_q^\mp {\rm J}_{q-s}(2D)
e^{-is\tau} \bar {\bf Y}^{(1)}
+\sum_s{\Delta\over s\pm E}A_s^\mp e^{-is\tau}\bar{\bf Y}^{(2)}
\nonumber\\
&&
=\sum_s A_s^\mp e^{-is\tau} e^{i2D\sin\tau}
\bar {\bf Y}^{(1)}
+\sum_s{\Delta\over s\pm E}A_s^\mp
e^{-is\tau}\bar{\bf Y}^{(2)}
\nonumber\\
&&
=\sum_s A_s^\mp e^{-is\tau} e^{i2D\sin\tau}
e^{-i D\sin \tau}{1\over\sqrt{2}}(1,-1)^t
+\sum_s{\Delta\over s\pm E}A_s^\mp e^{-is\tau}
e^{i D\sin \tau} {1\over\sqrt{2}}(1,1)^t
\nonumber\\
&&
=\sum_s A_s^\mp e^{-is\tau}{1\over\sqrt{2}}[
(1,-1)^t
+{\Delta\over s\pm E}(1,1)^t]e^{i D\sin \tau}
\nonumber\\
&&
=e^{i D\sin \tau}
\sum_s A_s^\mp e^{-is\tau}{1\over\sqrt{2}}
({\Delta\over s\pm E}+1,
{\Delta\over s\pm E}-1)^t,
\label{e14a}
\end{eqnarray}
where the superscript $\mp$ denotes the 
solutions corresponds to $\pm E$.

\section{Freeman Resonances}

In our solutions, we identify a Freeman resonance
when the new energy spacing $2E$, which is 
field-intensity dependent, is an integral multiple
of the field photon energy. The derived 
wave functions apparently have singularities only 
at $2E=2s\hbar\omega$ ($s=0,\pm1,\pm2,\pm3,...$), 
and not at the pre-resonance case. Here we do not 
exclude the $2E=0$ case, since $2E=0$ means $2E$ 
can be any integer. We also see from the wave function 
that in the pre-resonance case, the wave functions 
are finite, as well as the quasi energies given by 
eqs. (\ref{arccos}) and (\ref{arcsin}).

All this means that the pre-resonances are not
true resonances; only Freeman resonances are true 
resonances. At first glance, one may think that 
the Freeman resonances occur only when the 
resonating photon number is an even number. 
Since the quasi-energy spacing $2E$ can be added 
with an arbitrary integral multiple of the photon 
energy, we can replace $2E'=2E+1$ in above equations 
related to the wave functions. Thus, we immediately 
find the resonances with odd photon numbers. This 
analysis also indicates that we need four or more 
different quasi-energy levels as basic ones even 
in the non-resonance case, since $2E'=2E+1$ may
give different wave functions. On the other hand, 
we do not need more than four as basic ones, since 
$2E'=2E+2$ will not give a new wave function and 
it is included in the iteration process in the
continued fractions. Thus, we conclude that we 
need four and only four different quasi-energies 
as basic ones to produce the wave functions.

At the Freeman resonances, the intensity of the 
field and the original energy spacing $2\Delta$ 
satisfy the following equation from
eq. (\ref{cosshft}),
\begin{eqnarray}
\cos(2\pi\Delta)+2\pi R_{+} \sin(2\pi\Delta)=(-1)^n.
\label{freereson}
\end{eqnarray}
When $n=odd$, the equation reduces to
\begin{eqnarray}
2\pi R_{+} = -\cot^2(\pi\Delta).
\label{freeres1}
\end{eqnarray}
When $n=even$, the equation reduces to
\begin{eqnarray}
2\pi R_{+} = \tan^2(\pi\Delta).
\label{freeres2}
\end{eqnarray}
In this case, as we pointed before, the wave functions
have an infinite amplitude.

Equations (\ref{freeres1}) and (\ref{freeres2}) are 
transcendental equations. For a given field intensity 
and a Freeman resonance ($E=n\omega$), one can solve 
the transcendental equation to identify the resonating 
atomic level, which has the original spacing $2\Delta$ 
from the ground state. This equation has discrete 
solutions for $2\Delta$, which change when the field 
intensity changes. This theoretical feature does 
agree with experimental findings. Experimentalists 
call an electron energy peak in an ATI spectrum a 
Freeman resonance when the energy spacing between 
the ground state and the energy peak is an integral 
multiple of the laser photon energy, and intepret the 
energy peak as a formation of photoelectrons from 
a shifted Rydberg state. ATI spectra do show that the
Freeman resonances have a discrete feature for a fixed 
laser intensity. When the laser intensity changes, 
different sets of Rydberg levels come into play 
consecutively as Freeman resonances appearing in 
the ATI spectrum.

\section{Discussions}

The basic requirement to a correct solution of an interacting
system is when the interaction vanishes, the solution reduces to
the one of the corresponding non-interacting system. For the
problem in hand, the interaction is imposed through the dipole
moment, $D$. Thus the leading term of quasi-energy in the
expansion in powers of $D^2\omega^2$ should satisfy the basic
requirement, also should signify the physical meaning of field
intensity. Another important quantity is $(2\Delta-\omega)^2$
which signifies the detuning of the field frequency from the
transition frequency of the original two-level atom and competes
with $D^2\omega^2$ in the near pre-resonance processes. To see the
interesting competition in different limiting processes, we
consider the following two cases respectively.

\subsection{Energy Shift at Exact n-photon Pre-Resonances}

Practically, if we can tune up a
laser in a way that an integral multiple of
the laser frequency, $n\omega$, matches
the energy spacing, $2\Delta$, of a two level atom
i.e., $(2\Delta-n\omega)^2\ll D^2\omega^2$, we can
set $2\Delta=n\omega$ and use
eqs. (\ref{cosshft'}), (\ref{arccos}),
(\ref{arcsin}), (\ref{bsarccos}), and
(\ref{bsarcsin}) to obtain the energy shift in the
small-dipole limit.

We have the following three sub-cases:

1. The Single-Photon Case

In the single-photon pre-resonance case, the
quasi-energy $E$ has a simple form by just
keeping the leading term of $D$, with $\omega$ 
recovered in the expression explicitly,
\begin{eqnarray}
E={\omega\over \pi}\cos^{-1}
\left(\pm{\pi D \over 2}\right).
\label{e1pho}
\end{eqnarray}
Using $\cos^{-1}(x+d)=\cos^{-1}x-(1-x^2)^{-1/2}d$, 
where $x=0$ and $d=\pm\pi D/2$, we find
\begin{eqnarray}
E={\omega\over 2}(1+ D), \quad 
E_{BS}= \omega D. 
\label{e1pho'}
\end{eqnarray}
where the negative sign in eq. (\ref{e1pho}) is 
selected according to the limiting process from 
the near pre-resonance case in the next subsection.

This is a simple interesting results that the 
shift is proportional to $D$.

2. The Odd-Number Photon Case ($n\ne 1$)

\begin{eqnarray}
E={\omega\over \pi}\cos^{-1}
\left(\pm{\pi D^2 \over 2} {n\over n^2-1}\right). 
\label{eoddpho}
\end{eqnarray}
Using $\cos^{-1}(x+d)=\cos^{-1}x-(1-x^2)^{-1/2}d$, 
where $x=0$ and $d=\pm n\pi D^2/2(n^2-1)$, we find
\begin{eqnarray}
E={n\omega\over 2}(1+ {D^2\over n^2-1}),\quad 
E_{BS}= {n\omega D^2\over n^2-1}, 
\label{oddpho'}
\end{eqnarray}
with the same sign selection rule as above.

3. The Even-Number Photon Case

\begin{eqnarray}
E={\omega\over \pi}\sin^{-1}
\left(\pm{\pi D^2 \over 2} {n\over n^2-1}\right). 
\label{evenpho}
\end{eqnarray}
Using $\sin^{-1}(x+d)=\sin^{-1}x+(1-x^2)^{-1/2}d$, 
where $x=0$ and $d=\pm n\pi D^2/2(n^2-1)$, we find
\begin{eqnarray}
E={n\omega\over 2}(1+ {D^2\over n^2-1}),\quad 
E_{BS}= {n\omega D^2\over n^2-1}, 
\label{evenpho'}
\end{eqnarray}
where the positive sign is selected in eq. 
(\ref{evenpho}) according to the same limiting 
process as dsicussed above.

Now we see that the BS shift is proportional to 
$D$ in the single-photon pre-resonance case, 
while it is proportional to $D^2$ in the 
multiphoton pre-resonance case.

\subsection{Energy Shift in the Weak-Field and Near
Pre-Resonance Case }

In the previous case, we let $(2\Delta-n\omega)^2$
be infinitesimal first. In the rpesent case, 
switching the limiting procedures we let 
$D^2\omega^2$ be infinitesimal first, i.e.,
$D^2\omega^2 \ll (2\Delta-n\omega)^2$.

To treat this case, we need to expand
the term $2\pi \sin(2\pi\Delta) R_{+}$
in eq. (\ref{cosshft}) as
\begin{eqnarray}
&&
2\pi\sin(2\pi\Delta)\sum_{j=1}^{\infty}
({1\over2\Delta-j}+{1\over2\Delta+j})r_j
\nonumber\\
&&
\approx(-1)^n 2\pi^2
\{ r_n
+(2\Delta-n)(\sum_{j\ge1,\ne n}{1\over n-j}
+\sum_{j=1}^{\infty}{1\over n+j})r_j
\nonumber\\
&& -(2\Delta-n)^2{\pi^2\over6}r_n -(2\Delta-n)^2[\sum_{j\ge1,\ne
n}{1\over (n-j)^2} +\sum_{j=1}^{\infty}{1\over (n+j)^2}]r_j \}.
\label{cos02-3}
\end{eqnarray}

We consider the following sub-cases:

1. The Single-Photon Case

In the small-dipole case we only keep $r_1$, 
because only $r_1$ has $D^2$ as its leading 
order, while all other $r_n$ have $D^4$ as 
their leading order. Thus, from eqs. 
(\ref{cosshft}) and (\ref{laurent}), the 
cosine energy relations read
\begin{eqnarray}
\cos(2\pi {E\over\omega})
=-1+{\pi^2\over2}({2\Delta\over\omega}-1)^2
-2\pi^2r_1 -\pi^2 r_1({2\Delta\over\omega}-1), 
\label{i520}
\end{eqnarray}
where $r_1\approx -D^2/4$ in the leading order.
By expanding $\cos(2\pi E/\omega)
\approx -1 +\pi^2[(2E/\omega)-1]^2/2$,
we obtain an approximated quadratic equation
\begin{eqnarray}
(2E-\omega)^2= (2\Delta-\omega)^2
+D^2\omega^2+{D^2\over2}\omega(2\Delta-\omega),
\label{quadr}
\end{eqnarray}
which has solution
\begin{eqnarray}
E_{BS}\equiv (2E-2\Delta)
=-(2\Delta-\omega)\pm \sqrt{(2\Delta-\omega)^2
+D^2\omega^2+{D^2\over2}\omega(2\Delta-\omega)}
\label{ebsold}
\end{eqnarray}
This result agrees with the one from RWA but with 
an extra correction term.

For a small dipole $D^2\ll 1$, choosing the positive 
sign, the BS shift reduces to
\begin{eqnarray}
E_{BS}\approx{D^2\omega^2\over2(2\Delta-\omega)}
+{1\over4}D^2\omega. \label{ebsold1}
\end{eqnarray}
The extra term signifies the one-photon emission effect 
beyond the RWA. Here we see that when 
$ D^2\omega^2 \ll (2\Delta-\omega)^2$, i.e., the 
interaction term is smaller than the detuning, the BS
shift is proportional to $D^2$.

The sign selection here is forced by the requirement 
that when the interaction $D\omega \to 0$, the
value of $E_{BS}$ has to vanish. With the positive 
sign selected in eq. (\ref{ebsold}), letting 
$(2\Delta-\omega) \to 0$ directly, we get the same 
result as eq. (\ref{e1pho'}), whose sign is thus 
determined.

2. The Multiphoton Case

We have the following expansion:
\begin{eqnarray}
\cos(2\pi {E\over\omega})
=(-1)^n\{1-{\pi^2\over2}({2\Delta\over\omega}-n)^2\} 
+(-1)^n2\pi^2 ({1\over n-1}+{1\over n+1})
({2\Delta\over\omega}-n) r_1.
\label{i52t}
\end{eqnarray}

By expanding $\cos(2\pi E/\omega)
\approx (-1)^n\{1 -\pi^2[(2E/\omega)-n]^2/2\}$,
we obtain an approximated quadratic equation:
\begin{eqnarray}
(2E-n\omega)^2= (2\Delta-n\omega)^2 +D^2\omega{2n\over
n^2-1}(2\Delta-n\omega), \label{quadrt}
\end{eqnarray}
which has teh solution
\begin{eqnarray}
E_{BS}\equiv (2E-2\Delta) =D^2\omega{n\over n^2-1},
\label{ebsoldt}
\end{eqnarray}
where the sign is determined by the same method 
as before, which also justifies the sign selection
in eqs. (\ref{oddpho'}) and (\ref{evenpho'}).
An amazing thing here is that in the two different 
limiting processes and with different formulas,
$E_{BS}$ behaves the same way, both proportional 
to $D^2$ and with the same proportionality 
constant!

\section{Conclusions}

In the following we summarize the features of the
exact solution obtained:

1) It explicitly exhibits the Floquet quasi
energy behavior, namely the quasi-energy spacing 
is determined only up to its cosine.

2) It incorporates all multi-photon effects.

3) When the original energy spacing matches
an integer number of photon energy, the 
pre-resonance case, the interacting system
does not really resonate. Both quasi-energies
and wave functions are finite.

4) We have shown theoretically that when the 
shifted energy spacing matches an integer 
number of photon energy, the interacting 
system resonates. This theoretical feature 
explains the Freeman resonances observed in 
ATI experiments. We have also proven that the
Freeman resonances have a discrete feature, 
and when the field intensity changes different 
sets of unoccupied, excited atomic levels come 
into play as the resonances.

5) The Bloch-Siegert shift vanishes when the
interaction $D$ vanishes. The way it vanishes 
obeys some simple rules: Near the single-photon 
pre-resonance, whether the BS shift vanishes 
as $D$ or $D^2$ depends on whether the detuning 
first goes to zero or the interaction strength 
first goes to zero. Near the multiphoton 
pre-resonances, the BS shift always vanishes 
as $D^2$, independent of the limiting process. 
These rules can be subject to experimental tests.

{\it Acknowledgments} \hspace{0.3cm} DSG was 
supported in part by a summer grant of Physics 
Department of Ohio State University. YSW was 
supported in part by the NSF through grant 
PHY-9970701 and PHY-0457018.

\section{appendix}

We will show that $R_{+}$ as the residue of 
$\det(E)$ vanishes when $2\Delta=0$, and that
\begin{eqnarray}
R_{+}^{\infty}= \lim_{\Delta\to\infty} R_{+}
=\left|\matrix{ 1 &{D\over3\cdot2 } &\cdot  &\cdot
&\cdot &\cdot &\cdot \cr
{D\over2\cdot2 }& 1& {D\over2\cdot2 }
&\cdot &\cdot &\cdot &\cdot \cr
\cdot
& {D\over1\cdot2 }& 1& {D\over1\cdot2 }
&\cdot &\cdot &\cdot \cr
\cdot  &\cdot
& {D\over 2} & 0 &{D\over 2}
&\cdot &\cdot \cr
\cdot &\cdot &\cdot
& {D\over-1\cdot2 } & 1 & {D\over-1\cdot2 }
&\cdot \cr
\cdot  &\cdot &\cdot &\cdot
& {D\over-2\cdot2 }
& 1 &{D\over-2\cdot2 }\cr
\cdot &\cdot
&\cdot &\cdot &\cdot &{D\over-3\cdot2 } &1 \cr}
\right|=0.
\label{e46}
\end{eqnarray}

In the case of $2\Delta=0$,
\begin{eqnarray}
 \det (E)\vert_{\Delta=0} =
\left|\matrix{ 1 &{D\over E+3  } &\cdot  &\cdot &\cdot &\cdot
&\cdot  \cr {D\over E+2  } & 1&{D\over E+2 } &\cdot &\cdot &\cdot
&\cdot\cr \cdot & {D\over E+1  } & 1&{D\over E+1 } &\cdot &\cdot
&\cdot\cr \cdot  &\cdot & {D\over E } & 1 &{D\over E } &\cdot
&\cdot\cr \cdot &\cdot &\cdot & {D\over E-1 } & 1 &{D\over E-1 }
&\cdot \cr \cdot  &\cdot &\cdot &\cdot & {D \over E-2 } & 1
&{D\over E-2 }\cr \cdot &\cdot &\cdot &\cdot &\cdot &{D\over E-3 }
&1 \cr} \right|. \label{e48}
\end{eqnarray}
We notice
\begin{eqnarray}
\lim_{E\to n}(E-n)\det(E)\vert_{\Delta=0} =R_{+}(\Delta=0),
\label{e49}
\end{eqnarray}
which does not depend on $n$; thus,
\begin{eqnarray}
\det(E)\vert_{\Delta=0} =1+\pi R_{+}(\Delta=0)\cot(\pi E).
\label{e50}
\end{eqnarray}
The characteristic equation is obtained by
setting $\det(E)\vert_{\Delta=0}=0$, i.e.,
\begin{eqnarray}
\sin(\pi E)+\pi R_{+}(\Delta=0)\cos(\pi E) =0, \label{e51}
\end{eqnarray}
which has solutions
\begin{eqnarray}
E = -{1\over\pi}\tan^{-1}(\pi R_{+}(\Delta=0)), \label{e52}
\end{eqnarray}
where $R_{+}(\Delta=0)$ is
\begin{eqnarray}
R_{+}(\Delta=0) =\left|\matrix{ 1 &{D\over3 } &\cdot  &\cdot
&\cdot &\cdot &\cdot \cr {D\over2 }& 1& {D\over2 } &\cdot &\cdot
&\cdot &\cdot \cr \cdot & D& 1& D &\cdot &\cdot &\cdot \cr \cdot
&\cdot & {D\over 2} & 0 &{D\over 2} &\cdot &\cdot \cr \cdot &\cdot
&\cdot & -{D\over1} & 1 & -{D\over1} &\cdot \cr \cdot  &\cdot
&\cdot &\cdot & -{D\over2 } & 1 &-{D\over2}\cr \cdot &\cdot &\cdot
&\cdot &\cdot &-{D\over3 } &1 \cr} \right|=0. \label{e53}
\end{eqnarray}
The last step ( $=0$ ) needs a proof.

{\it Proof}:

We use a recurrence relation of Bessel functions
\begin{eqnarray}
-{x\over 2n}(J_{n-1}+ J_{n+1})+J_{n}=0,\nonumber\\
 K(J_{-1}+ J_{1})=0,
\label{e54}
\end{eqnarray}
where $K$ is an arbitrary constant.

Let $n$ run from $-\infty$ to $\infty$. The algebraic equation set
for $J_n$ has non-zero solutions. So the coefficient determinant
must be zero, i.e.,
\begin{eqnarray}
\det(x)=\left|\matrix{ 1 &{-x\over6 } &\cdot  &\cdot
&\cdot &\cdot &\cdot \cr
{-x\over4 }& 1& {-x\over4 }
&\cdot &\cdot &\cdot &\cdot \cr
\cdot
& {-x\over2 }& 1& {-x\over2 }
&\cdot &\cdot &\cdot \cr
\cdot  &\cdot
& {K} & 0 &{{K}}
&\cdot &\cdot \cr
\cdot &\cdot &\cdot
& {x\over2} & 1 & {x\over2}
&\cdot \cr
\cdot  &\cdot &\cdot &\cdot
& {x\over4 }
& 1 &{x\over4 }\cr
\cdot &\cdot
&\cdot &\cdot &\cdot &{{x\over6 } } &1 \cr}
\right|=0.
\label{e55}
\end{eqnarray}
By setting $x=-D$ and $K=D/2$ we get eq. (\ref{e46}).
By setting $x=-2D$ and $K=D/2$ we get eq. (\ref{e53}).

{\it QED}


\end{document}